\documentclass[12pt]{ronbun}

\usepackage{amsmath,amssymb}
\usepackage{graphicx,color}
\usepackage{cite}
\usepackage{bm}
\usepackage{dcolumn}

\newcommand{\nn}{\nonumber}
\newcommand{\e}{{\rm e}}

\newcommand{\del}{\delta}
\newcommand{\ra}{\rangle}
\newcommand{\la}{\langle}

\newcommand{\al}{\alpha}

\renewcommand{\th}{\theta}

\numberwithin{equation}{section}
\begin{document}

\begin{flushright}
\parbox{4.2cm}
{KEK-TH-990 \hfill \\
UTHEP-499 \\
{\tt hep-th/0410139}
 }
\end{flushright}

\vspace*{0.7cm}

\begin{center}
 \Large\bf 
Open Semiclassical Strings and Long Defect Operators 
in AdS/dCFT Correspondence 
\end{center}
\vspace*{0.7cm}
\centerline{\large Yoshiaki Susaki$^{\ast\dagger a}$, 
Yastoshi Takayama$^{\ddagger b}$ 
and Kentaroh Yoshida$^{\ast c}$}
\begin{center}
$^{\ast}$\emph{Theory Division,   	
Institute of Particle and Nuclear Studies, \\
High Energy Accelerator Research 
Organization (KEK),\\
Tsukuba, Ibaraki 305-0801, Japan.} 
\\
$^{\dagger}$\emph{Institute of Physics, University of Tsukuba, \\
Tsukuba, Ibaraki 305-8571, Japan.} 
\\
$^{\ddagger}$\emph{
Department of Particle and Nuclear Physics, \\
The Graduate University for Advanced Studies, \\
Tsukuba, Ibaraki 305-0801, Japan.} 

\vspace*{0.3cm}
$^a${\tt susaki@post.kek.jp} \quad  
$^b${\tt takaya@post.kek.jp} \quad 
$^c${\tt kyoshida@post.kek.jp}
\end{center}

\vspace*{0.7cm}

\centerline{\bf Abstract} 

We consider defect composite operators in a defect superconformal field
theory obtained by inserting an AdS$_4\times$S$^2$-brane in the
AdS$_5\times$S$^5$ background. The one-loop dilatation operator for the
scalar sector is represented by an integrable open spin chain.  We give
a description to construct coherent states for the open spin
chain. Then, by evaluating the expectation value of the Hamiltonian with
the coherent states in a long operator limit, a Landau-Lifshitz type of
sigma model action is obtained. This action is also derived from the
string action and hence we find a complete agreement in both SYM and
string sides. We see that an $SO(3)_{\rm H}$ pulsating string solution
is included in the action and its energy completely agrees with the
result calculated in a different method. In addition, we argue that our
procedure would be applicable to other AdS-brane cases.

\vspace*{0.5cm}

\vfill \noindent {\bf Keywords:}~~{\footnotesize AdS/CFT, spin chain,
integrability, semiclassical string, defect CFT}

\thispagestyle{empty}
\setcounter{page}{0}

\newpage

\section{Introduction}

One of the most important subjects in string theory is to test the
AdS/CFT correspondence \cite{M} beyond BPS sectors.  In the analysis at
almost BPS region, a great development was made by
Berenstein-Maldacena-Nastase (BMN) \cite{BMN}. They presented the
AdS/CFT correspondence at string theoretical level by using the Penrose
limit \cite{P,BFHP1} and the exact solvability of pp-wave string theory
\cite{MT}. The pp-wave string states and their energies correspond to
the BMN operators and the full conformal dimensions (including the
anomalous dimensions) in $\mathcal{N}$=4 super Yang-Mills (SYM),
respectively.  

The BMN analysis was generalized by considering the semiclassical
quantization \cite{GKP2,FT} around classical rotating string solutions,
instead of pp-wave strings. The semiclassical string states and their
energies correspond to certain single-trace local operators and their
conformal dimensions in SYM in specific regimes.\footnote{The classical
integrability of type IIB string on the AdS$_5\times$S$^5$ background
\cite{BPR} may be deeply related to the correspondence between string
and SYM sides \cite{DNW}. The matching of the spectra and the
equivalence of the integrable structures between the spin chain and
string theories are confirmed up to and including the two-loop order on
a specific example \cite{AS}. } The energy of the semiclassical string
state can be expanded in positive power of new effective coupling
$\tilde{\lambda}\equiv \lambda/J^2$ ($\lambda$ is 't Hooft coupling)
when the classical spinning string solutions have, at least, one of the
S$^5$-spins $J_1,~J_2$ and $J_3$\,. It is therefore possible to compare
the energy with the conformal dimension perturbatively calculated in
$\mathcal{N}$=4 SYM. Several types of classical solutions which give
regular expansions of energies have been discovered by some
works\cite{GKP2,FT,FT2,ART} (For a review, see \cite{review}). On the
other hand, the anomalous dimensions of the single-trace composite
operators can be computed by using the Bethe ansatz techniques. This
fact was firstly shown by Minahan and Zarembo in a study of the $SO(6)$
scalar sector \cite{MZ}. This kind of analysis is applicable to other
sectors \cite{BKS} such as $SU(2)$ \cite{BFST,BMSZ,Kazakov}, $SU(3)$
\cite{SU(3)}, $SL(2)$ \cite{BFST,SL2,QCD,BS,KZ}\footnote{The
integrability related to the $SL(2)$ Heisenberg spin chain was
previously found in non-supersymmetric QCD. For example, see
\cite{QCD}.}, $SU(2|3)$ \cite{su(2|3)} and the full $PSU(2,2|4)$ sectors
\cite{BS}.  In particular, the $SU(2)$ sector is well studied.  An
approach to the analysis with the Inozemtsev long range spin chain at
three-loop level \cite{SS} and, after that, the asymptotic 
all-loop Bethe ansatz\footnote{This is indeed asymptotic in
a sense that it is unsuitable for describing the operators of small
dimensions at sufficiently high orders of perturbation theory
(e.g.\,certain interactions of wrapping type are missed). } 
is proposed \cite{BDS}. 

For the Bethe ansatz, in parallel to studies of quantum strings on the
AdS$_5\times$S$^5$\,, new types of Bethe ansatz for them were developed
\cite{quantum}. This important development should be a promising
procedure to investigate the quantum aspects of the AdS-string.
\vspace*{0.3cm}

In this paper we study the correspondence between semiclassical strings
and long scalar operators in open string cases.  By inserting spatial
defects (called AdS-branes) in the bulk AdS$_5\times$S$^5$ background
\cite{KR}, one may consider open strings in the context of the AdS/CFT
correspondence where the CFT side has defects and so becomes a
defect conformal field theory (dCFT).  We focus on the case that an
AdS$_4\times$S$^2$-brane is inserted.  The dCFT was particularly
investigated by DeWolfe-Freedman-Ooguri \cite{DFO} and it was shown to
be superconformal \cite{EGK}. The BMN operator correspondence for this
system was also studied by Lee and Park \cite{LP}. Furthermore, the
matrix of one-loop anomalous dimension for defect operators (which have
the defect fields in the end-points, instead of the trace) in the scalar
sector was calculated by DeWolfe and Mann \cite{DM}\footnote{For open
strings in other setup, see \cite{BGMNN,BHLN} and
\cite{Open-sol,CWW}.}. The resulting dilatation operator is an
integrable {\it open} spin chain Hamiltonian while a {\it closed} spin
chain appears in the case of single trace operators. By using the Bethe
ansatz techniques, the exact BMN operators \cite{B} for the open strings
were derived as in \cite{MZ}. In the long operator limit, those are
consistently reduced to the result of \cite{LP}. On the other hand,
except the BMN operators, the string side is not uncovered yet. 

Motivated by this fact, we will reveal the open semiclassical string
corresponding to the open spin chain by using the method 
of \cite{Kruczenski,KRT} with coherent state \cite{coherent}. 
We give a description to construct coherent
states for the defect operators and then the expectation value of the
open spin chain Hamiltonian is taken with them. In a continuum limit
(long operator limit), the Landau-Lifshitz (LL) type sigma model action
is obtained as in \cite{ST-coh,KT-coh,review-coh}. Namely, the first
time-derivative appears instead of the usual quadratic one. On the other
hand, we derive the sigma model from the string action with a suitable
gauge-fixing \cite{KT-coh,review-coh}. In conclusion, we find a complete
agreement of the sigma model actions as in
\cite{Kruczenski,KRT,HL,ST-coh,KT-coh,review-coh}. We also see that the
resulting sigma model action includes a pulsating solution and its
energy is recovered by putting the solution ansatz into the action.

\vspace*{0.5cm}
This paper is organized as follows: In section 2, the dCFT we consider
is briefly introduced. Section 3 is devoted to a short review of the
one-loop result in the scalar sector \cite{DM}. In section 4, 
we give a description to construct
coherent states for the open spin chain Hamiltonian. Then we take the
expectation value of the Hamiltonian, and the LL type sigma
model is obtained in a continuum limit. We find that an $SO(3)_{\rm H}$
pulsating string solution is included in this action. In section 5, we
rederive the sigma model obtained in section 4 from the string
action. In section 6, we discuss coherent states for other AdS-brane
cases. One may see that our consideration for an AdS$_4\times$S$^2$
brane would be extended to other AdS-branes.  Section 7 is devoted to a
conclusion and discussions.

\section{Setup of Defect Conformal Field Theory} 

\begin{figure}[htbp]
 \begin{center}
  \includegraphics[scale=.7]{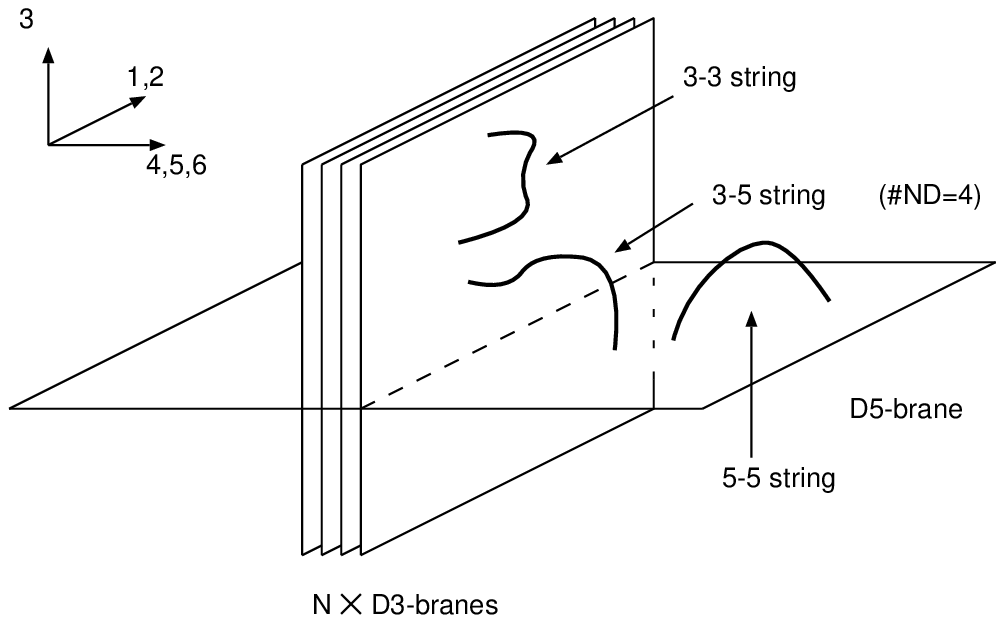}
 \end{center}
\vspace*{-0.5cm}
\caption{An intersecting D5-brane with a stack of $N$ D3-branes.}
\label{D3-D5:fig}
\end{figure}

From now on, we will consider a dCFT by inserting a supersymmetric
AdS$_4\times$S$^2$ brane into the AdS$_5\times$S$^5$ background
\cite{DFO}. The AdS$_4$ brane inside AdS$_5$ could be naturally realized
in string theory by considering a supersymmetric D5-brane intersecting a
stack of $N$ D3-branes as depicted in Fig.\,\ref{D3-D5:fig}. The
near-horizon limit of the $N$ D3-branes produces an AdS$_5\times$S$^5$
background where the D5-brane is realized as an AdS$_4\times$S$^2$
submanifold. That is, in this system, closed strings propagating
throughout the spacetime provide a holographic description of an
$\mathcal{N}$=4 SYM$_4$ on the boundary of AdS$_5$\,. In addition, the
fluctuations on the AdS$_4$-brane should be dual to additional physics
confined to the boundary of the AdS$_4$\,, and the dual field theory
contains new fields living on a $(1+2)$-dimensional defect, obtained
from the low-energy limit of the 3-5 open strings interacting with the
3-3 strings of the original brane setup. The 5-5 strings are realized as
open strings on the AdS$_4\times$S$^2$ brane in the gravity side.

\vspace*{0.2cm}

In the CFT side, $\mathcal{N}$=4 SYM has a three-dimensional defect
due to the presence of the AdS-brane. One may see the usual
$\mathcal{N}$=4 SYM in the long distance from the defect, while
$\mathcal{N}$=4 SYM couples to the defect fields in the neighborhood of
the defect. Then the vector multiplet of $\mathcal{N}$=4 in four
dimensions decomposes into a vector multiplet and a hypermultiplet in
three dimensions. 

Now let us introduce the field contents of $\mathcal{N}$=4 SYM and
dCFT. The $\mathcal{N}$=4 vector multiplet is composed of a gauge field
$A_{\mu}~(\mu=0,1,2,3)$, adjoint Majorana spinors
$\lambda^{\al}~(\al=1,2,3,4)$ and six real scalars
$X^i~(i=1,\ldots,6)$\,. The $\lambda^{\al}$ and $X^i$ are in the 
${\bf 4}$ and ${\bf 6}$ of the $SO(6)$ R-symmetry, respectively. Our
convention basically follows that of \cite{DM}.  Then the defect
introduces an additional 3D hypermultiplet propagating on the
hypersurface $x^3 = 0$\,. This multiplet is composed of a complex scalar
$q^m$ and a complex 2-component fermion $\Psi^a$\,. These new fields
break the total symmetry, and the breaking pattern is as follows:
\begin{eqnarray}
&& \mbox{Conformal symmetry}: \quad SO(2,4) \rightarrow SO(2,3) \cong 
Sp(4)\,, \nn \\
&& \mbox{R-symmetry}: \quad 
SO(6)_{\rm R} \rightarrow SO(3)_{\rm H} \times SO(3)_{\rm V}\cong 
O(4)\,, \nn \\
&& \mbox{Superconformal symmetry}: \quad PSU(2,2|4) \rightarrow
 OSp(4|4)\,. \nn 
\end{eqnarray}
The defect fields 
$q^m$ and $\Psi^a$ transform 
in the $({\bf 2}, {\bf 1})$ and in the $({\bf 1}, {\bf 2})$ 
of $SO(3)_{\rm H}\times SO(3)_{\rm V}$\,, respectively. 
The $\mathcal{N}$=4 vector multiplet is also decomposed into 
\begin{eqnarray}
&& \mbox{3D vector multiplet}:~~
\{A_k, P_+\lambda^{\al}, X_{\rm V}^A, D_3X_{\rm H}^I\}\,, \nn \\ 
&& \mbox{3D hypermultiplet}:~~
\{A_3, P_-\lambda^{\al}, X_{\rm H}^I, D_3 X_{\rm V}^A\}\,, \nn 
\end{eqnarray}
where $k=0,1,2$ and $i=1,\ldots,6$ of $SO(6)$ is decomposed into
$A=1,2,3$ of ${\bf 3}$ in the $SO(3)_{\rm V}$ and $I = 4,5,6$ of ${\bf
3}$ in the $SO(3)_{\rm H}$\,. That is, $X_{\rm H}^I$ and $X_{\rm V}^A$
are in the $({\bf 3}, {\bf 1})$ and $({\bf 1}, {\bf 3})$ of the
$SO(3)_{\rm H}\times SO(3)_{\rm V}$\,, respectively. 

These defects couple directly only to the bulk vector multiplet, and
their dynamics is described by the following action\footnote{We will
consider the case that a single D5-brane intersecting with a stack of
$N$ D3-branes in this paper.  For $M \ll N$\,, we may consider the case
that $M$ multiple D5-branes intersect with $N$ D3-branes, and flavor
indices appear on $q$ and $\Psi$ in an obvious way, while the $M\sim N$
case is more involved and difficult because the backreaction of the
D5-branes should be considered.}
\begin{eqnarray}
S_3 &=& S_{\rm kin} + S_{\rm yuk} + S_{\rm pot}\,, \\
S_{\rm kin} &=& \frac{1}{g^2}\int\!\!d^3x\Bigl[
(D^k q^m)^{\dagger}D_k q^m -i\bar{\Psi}^a\rho^kD_k\Psi^a
\Bigr]
\,, \nn \\
S_{\rm yuk} &=& \frac{1}{g^2}\int\!\!d^3x\Bigl[
i\bar{\Psi}^aP_+\lambda_{am}q^m -i\bar{q}^m\bar{\lambda}_{ma}P_+\Psi^a 
+ \bar{\Psi}^a\sigma^A_{ab}X_{\rm V}^{A}\Psi^a
\Bigr]\,, \nn \\
S_{\rm pot} &=& \frac{1}{g^2}\int\!\!d^3x\Bigl[
\bar{q}^m X_{\rm V}^AX_{\rm V}^Aq^m 
+ i\epsilon_{IJK}\bar{q}^m\sigma_{mn}^{I}X_{\rm H}^JX_{\rm H}^Kq^n
\Bigr] \nn \\
&& + \frac{1}{g^2}\int\!\!d^3x\left[
\bar{q}^m\sigma^I_{mn}(D_3 X_{\rm H}^I)q^n 
+ \frac{1}{4}\delta(0){\rm Tr}(\bar{q}^m\sigma^I_{mn}q^n)^2
\right]\,, \nn
\end{eqnarray}
where the covariant derivative is defined as $D_k \ast \equiv \partial_k
\ast -i A_k \ast$\,. The total action of the theory is given by the
action of $\mathcal{N}$=4 SYM$_4$ and the above action for the defect
fields. Notably, the total theory has only one parameter, coupling
$g_{\rm YM}$\,, and so it is exactly marginal \cite{EGK}.

\section{The One-Loop Anomalous Dimensions of Defect Operators}

Here we shall briefly review the one-loop result obtained by DeWolfe and
Mann \cite{DM}. We are interested in open strings on the
AdS$_4\times$S$^2$-brane. Their states are described by local,
gauge-invariant defect operators composed of Lorentz scalar fields:
\begin{eqnarray}
\mathcal{O} = \psi_{m,j_1,\ldots,j_L,n}\,\bar{q}_mX^{j_1}\cdots
 X^{j_L}q^n\,. 
\label{cop}
\end{eqnarray}
Here two defect scalar fields appear at both ends of a sequence of
scalar fields,   
instead of the trace operation in closed string cases.  
By evaluating the correlation function,
\begin{eqnarray}
\la \bar{q}_{n'}(z_{\beta})X^{i_L}(z_L)\cdots X^{i_1}(z_1)
q^{m'}(z_{\al})\mathcal{O}(0)\ra\,, 
\end{eqnarray}
one can compute the matrix of anomalous dimension for (\ref{cop}).  In
the one-loop level analysis, the bulk interactions lead to the same
result in closed string cases, but the defect interactions give
additional terms to the matrix of anomalous dimension.

The matrix of anomalous dimension at one-loop level is 
computed by evaluating the one-loop planar diagrams. 
It is represented by the Hamiltonian of an integrable
open spin chain and it is composed of the bulk part and the defect
part as follows:
\begin{eqnarray}
&& D_{SO(6)} = \Gamma^{\rm bulk}_{\mathcal{O}} + 
\Gamma^{\rm defect}_{\mathcal{O}} =   \frac{\lambda}{16\pi^2}
\sum_{l=1}^{L-1}H_{l,l+1} + \frac{\lambda}{16\pi^2}\Bigl[
(2I_{\bar{\al}1} +2\bar{S}_{\bar{\al}1}) 
+ (2I_{L\beta} +2 S_{L\beta})
\Bigr]\,, 
\label{open-spin} \\
&& H_{l,l+1} = K_{l,l+1} + 2I_{l,l+1} -2 P_{l,l+1}\,. 
\end{eqnarray}
The bulk part (the first term in (\ref{open-spin})) is the same 
as the result of Minahan and Zarembo \cite{MZ}, and it is written by 
the trace $I$, permutation $P$ and trace $K$ operators: 
\begin{eqnarray}
I_{j_{l}j_{l+1}}^{i_{l}i_{l+1}} 
= \delta^{i_l}_{j_l}\delta^{i_{l+1}}_{j_{l+1}}\,, \quad 
P_{j_{l}j_{l+1}}^{i_li_{l+1}} = \delta^{i_l}_{j_{l+1}}
\delta_{j_l}^{i_{l+1}}\,, \quad 
K_{j_lj_{l+1}}^{i_li_{l+1}} = \delta^{i_li_{l+1}}\delta_{j_lj_{l+1}}\,.  
\end{eqnarray}
The defect part (the second term in (\ref{open-spin})) 
is written in terms of 
\begin{eqnarray}
&& I^{\bar{m} I}_{\bar{n} J} = \del^{\bar{m}}_{\bar{n}} 
\del_{J}^{I}\,, \quad 
I^{I m}_{J n} = \del_{J}^{I}\del^{m}_{n}\,, 
\quad I^{\bar{m} A}_{\bar{n} B} 
= I^{\bar{m} I}_{\bar{n} B} = I^{\bar{m} A}_{\bar{n} J} = 0\,, \quad 
I_{A m}^{B n} = I_{A m}^{J n} = I_{I m}^{B n} = 0\,, \nn \\  
&& S_{Jn}^{Im} = -i\epsilon_{IJK}\sigma^K_{mn}\,, \quad 
S_{Bn}^{Am} = \del^m_n\del^A_B\,, \quad S^{Im}_{Bn} = S^{Am}_{Jn} = 0\,, 
 \\ && 
\bar{S}_{\bar{n}J}^{\bar{m}I} = i\epsilon_{IJK}\sigma^K_{\bar{m}\bar{n}}\,, 
\quad~~
\bar{S}_{\bar{n}B}^{\bar{m}A} = \del_{\bar{n}}^{\bar{m}}\del^A_B\,, 
\quad \bar{S}^{\bar{m}I}_{\bar{n}B} = \bar{S}^{\bar{m}A}_{\bar{n}J} = 0\,. 
\nn 
\end{eqnarray} 
Here we should note that the bulk part has no periodicity. In the
present case, the integrable boundary arising from the defect
contribution ensures the integrability of the spin chain, instead of
periodicity.

It may be helpful for a check of calculations to use the chiral primary
operators \cite{DFO}, 
\begin{eqnarray}
\bar{q}^m \sigma_{mn}^{(I_1}X_{\rm H}^{I_2}X_{\rm H}^{I_3} \cdots 
X_{\rm H}^{I_{L+1})}q^n\,,
\label{chiral}
\end{eqnarray}
where parentheses denote total symmetrization and traceless with respect
to the index of $SO(3)_{\rm H}$\,. One can indeed verify that 
each $H_{l,l+1}$, as well as $2I_{\bar{\al}1} + 2\bar{S}_{\bar{\al}1}$
and $2I_{L\beta} + 2S_{L\beta}$\,, separately gives zero by acting them
to (\ref{chiral}). 

Finally, by introducing the $SO(6)$ generators $M_{ab}^{ij} \equiv
\del^{i}_a \del^j_b - \del_a^j\del_b^i$\,, a piece of the spin chain
Hamiltonian, $H_{l,l+1}$ is rewritten as 
\begin{eqnarray}
\label{int}
H_{l,l+1} = M_l^{ij}M_{l+1}^{ij} - \frac{1}{16}(M_l^{ij}M_{l+1}^{ij})^2 + \frac{9}{4}\,. 
\end{eqnarray}
In this form the spin-spin interaction is manifest. The expression
(\ref{int}) is useful to evaluate the expectation value of the spin
chain Hamiltonian with coherent states. 

\section{Coherent States and Sigma Model for Defect Operators}

\subsection{Construction of Coherent States for Open Spin Chain}

We will consider how to construct the coherent states for the open spin
chain. The open spin chain has boundaries which break the $SO(6)$
symmetry to $SO(3)_{\rm H} \times SO(3)_{\rm V}$\,. This fact reflects
that a D5-brane, whose shape is an AdS$_4\times$S$^2$ geometry, is
inserted into the AdS$_5\times$S$^5$\,.  We may consider various open
semiclassical strings attaching to the S$^2$ part, but we concentrate on
open strings sticking to the S$^2$ described in the three-dimensional
space spanned by $X_{\rm H}^I~(I=1,2,3)$ with the condition
$\sum_{I=1}^3(X_{\rm H}^I)^2 = 1$\,. It would be difficult to obtain a
regular semiclassical string energy (i.e., a regular BMN limit) in the
other cases. In fact, we have not succeeded so far.

Hence let us consider the following ansatz for the whole coherent state:
\begin{eqnarray}
 | m \ra = |\bar{q} \ra \otimes 
\prod_{l=1}^{L}|m_l\ra \otimes  |q \ra\,,  
\label{whole}
\end{eqnarray}
where $|m_l\ra $ are $SO(3)_{\rm H}$ coherent states, and $|\bar{q}\ra$,
$|q\ra$ are $SU(2)_{\rm H}$ ones. We need to construct and adjust the
these coherent states so that the well-defined continuum limit can be
taken.

For the $SO(3)_{\rm H}$ sector, the coherent states are described by
using a coset $SO(3)_{\rm H}/SO(2)$\,. Then, according to the choices of
the $SO(6)$ coherent states, there are two possibilities:
\begin{enumerate}
\item the vacuum $|0\ra = (0,1,i)$ and $H= SO(2)$\, where the vacuum is
invariant up to a phase factor. The Cartan generator is $M_{56}$ and the
coherent state is generated by $M_{45}$ and $M_{64}$\,. 
\item the vacuum $|0\ra =(0,0,1)$ and $H = SO(2)$\,. The Cartan
generator is $M_{45}$ and the coherent state is generated by $M_{56}$
and $M_{64}$\,.
\end{enumerate} 
Here we have extracted the vacuum and the generators for $SO(3)_{\rm
H}/SO(2)$ from the full $SO(6)$ case. As we will see below, the
``local'' BPS property (case 1) and the large ``extensive'' one-loop
shift of the dimension (case 2) are also inherited even after the
$SO(6)$ is spontaneously broken. We will concentrate on the case 1. In
the end of this section we will comment on the case 2. We will explain
the construction of $SO(3)_{\rm H}/SO(2)$ and $SU(2)_{\rm H}/U(1)$
coherent states below.

\subsubsection*{Construction of $SO(3)_{\rm H}/SO(2)$ coherent states}

Let us introduce $M^{IJ}$ as $SO(3)_{\rm H}$ generators in the
fundamental representation $(M^{IJ})_{ab} = \delta^I_a \delta^J_b -
\delta^I_b \delta^J_a ~(a,b=1,2,3,~I,J=1,2,3)$\footnote{We have
redefined the expressions of the $SO(3)_{\rm H}$ generators as $M_{45}$,
$M_{56}$, $M_{64}$ $\rightarrow$ $M_{31}$, $M_{12}$, $M_{23}$.}.
Choosing a vector $(1,i,0)$ as a vacuum $|0\ra$ (i.e., the highest
weight state for $M^+= M^{23}+iM^{31}$)\,, the $SO(3)_{\rm H}/SO(2)$
coherent state is given by
\begin{eqnarray}
 | m \rangle = \mbox{\boldmath{$m$}}
\equiv
 \frac{1}{\sqrt{2}}\exp[aM^{23}+bM^{31}]
	\left[
   \begin{array}{@{\,}c@{\,}}
    1\\
	 i\\
    0
	\end{array}
   \right]
= \frac{\e^{-i\phi}}{\sqrt{2}}
 \left[
	\begin{array}{@{\,}c@{\,}}
	\cos\theta \cos\phi + i \sin\phi\\
	-\cos\theta \sin\phi + i \cos\phi\\
	\sin\theta
	\end{array}
 \right]\,, \label{SO3m}
\end{eqnarray} 
where $\theta=\sqrt{a^2+b^2}$, $a=\theta\sin\phi$ and
$b=\theta\cos\phi$\,. One can explicitly check that $\langle m|m \rangle
= |\mbox{\boldmath{$m$}}|^2= 1\,, \quad \mbox{\boldmath{$m$}}^2 =
(\mbox{\boldmath{$m$}}^*)^2 = 0$\,. By putting this coherent state at
each $l$-th site as $|m_l\ra$\,,
\begin{eqnarray}
 |m_l \rangle = \mbox{\boldmath $m$}_l =
 \frac{\e^{-i\phi_l}}{\sqrt{2}}
 \left[
	\begin{array}{@{\,}c@{\,}}
	\cos\theta_l \cos\phi_l + i \sin\phi_l\\
	-\cos\theta_l \sin\phi_l + i \cos\phi_l\\
	\sin\theta_l
	\end{array}
\right]
\qquad (l=1,\ldots, L)\,, \label{m-coh}
\end{eqnarray}
the $SO(3)_{\rm H}$ coherent states for the
bulk part of the spin chain are constructed.

\subsubsection*{Construction of $SU(2)_{\rm H}/U(1)$ coherent states }

We have to determine two more coherent states $|\bar{q}\rangle$ and
$|q\rangle$\,. The $SU(2)_{\rm H}/U(1)$ coherent states can be written,
up to a $U(1)$ factor, as
\begin{eqnarray}
&& \hspace*{-0.5cm}
\bigg| \frac{SU(2)_{\rm H}}{U(1)}\bigg \rangle
=  \e^{i\alpha \sigma_1 + i\beta \sigma_2}
 \left[
	\begin{array}{@{\,}c@{\,}}
	1\\
	0
	\end{array}
 \right]
=
 \left[
	\begin{array}{@{\,}c@{\,}}
	\cos(\vartheta/2)\\
	\e^{-i \varphi} \sin(\vartheta/2)
	\end{array}
 \right] 
\simeq
 \left[
	\begin{array}{@{\,}c@{\,}}
	\e^{i \varphi /2}\cos(\vartheta/2)\\
	\e^{-i \varphi /2} \sin(\vartheta/2)
	\end{array}
 \right]\,, 
\end{eqnarray}
where $\vartheta = 2\sqrt{\alpha^2+\beta^2}\,,~
\alpha = \frac{\vartheta}{2}\sin\varphi\,,~\beta 
= \frac{\vartheta}{2}\cos\varphi$\,. 
The vacuum $(1,0)$ is the highest weight state
for $\sigma_+=\sigma_1+i\sigma_2$\,. We take 
$|\bar{q}\rangle$ and $|q\rangle$ as
\begin{eqnarray}
 |\bar{q}\ra = \left[
	\begin{array}{@{\,}c@{\,}}
	\e^{-i \varphi_{\bar q} /2}\cos(\vartheta_{\bar q}/2) \\
	\e^{i \varphi_{\bar q} /2} \sin(\vartheta_{\bar q}/2)
	\end{array}
 \right]\,, \qquad
 |q \rangle = \left[
	\begin{array}{@{\,}c@{\,}}
	\e^{i \varphi_q /2}\cos(\vartheta_q/2)\\
	\e^{-i \varphi_q /2} \sin(\vartheta_q/2)
	\end{array}
 \right]\,.
\label{q-coh}
\end{eqnarray} 
These coherent states satisfy 
\begin{eqnarray}
\la \bar{q} | \mbox{\boldmath{$\sigma$}}  | \bar{q} \ra =
\la q | \mbox{\boldmath{$\sigma$}}^T  | q \ra 
= \bigl(
\sin\vartheta \cos\varphi, \sin\vartheta \sin \varphi, \cos\vartheta \bigr) 
\equiv \mbox{\boldmath $n$}\,.
\end{eqnarray} 

We next discuss the expectation values of $\Gamma_{\cal O}^{\rm
defect}$\,. The boundary parts are described as
\begin{eqnarray}
 |{\cal L} \ra = | m_1 \ra \otimes |\bar{q} \ra\,, \qquad |{\cal R} \ra
 = | m_L \ra \otimes |q \ra\,.
\end{eqnarray} 
For the left end-point, the expectation value is\footnote{For a moment,
we omit the factor $\frac{\lambda}{8\pi^2}$ for convenience since it is
irrelevant here.}
\begin{eqnarray}
 \langle {\cal L} | \Gamma^{\rm defect-L}_{\cal O} | {\cal L} \ra &\simeq&
 \langle {\cal L} | I_{\bar{q}1}+S_{\bar{q}1} | {\cal L} \ra 
= 1 + i\mbox{\boldmath $n$}_{\bar{q}}\cdot(\mbox{\boldmath $m$}_1^{\ast}\times 
\mbox{\boldmath $m$}_1) \nn \\
&=& 1 -(\sin\vartheta_{\bar{q}} \cos \varphi_{\bar q}, 
\sin\vartheta_{\bar{q}} \sin \varphi_{\bar q}, \cos\vartheta_{\bar q})
\left[
\begin{array}{@{\,}c@{\,}}
\sin \theta_1 \cos(\pi-\phi_1)\\
\sin \theta_1 \sin(\pi-\phi_1)\\
\cos\theta_1
\end{array}
\right]\,.
\end{eqnarray} 
In a similar way, for the right end-point, we obtain 
\begin{eqnarray}
 \langle {\cal R} | \Gamma^{\rm defect-R}_{\cal O} | {\cal R} \ra 
 &\simeq& \langle {\cal R} | I_{Lq}+S_{Lq} | {\cal R} \ra 
=  1 - i\mbox{\boldmath $n$}_{q}\cdot(\mbox{\boldmath $m$}_L^{\ast}\times 
\mbox{\boldmath $m$}_L) \nn \\ 
&=& 1 + (\sin\vartheta_{q} \cos \varphi_{q},
 \sin\vartheta_{q} \sin \varphi_{q}, \cos\vartheta_{q}) \left[
	\begin{array}{@{\,}c@{\,}}
	\sin \theta_L \cos(\pi-\phi_L)\\
	\sin \theta_L \sin(\pi-\phi_L)\\
	\cos\theta_L
	\end{array}
\right]\,.
\end{eqnarray} 

Now let us consider the following conditions:
\begin{eqnarray}
 \left\{
 \begin{array}{c} 
  \displaystyle 
\vartheta_{\bar q} =  \theta_1 + \frac{1}{L}\delta\theta_1 
+ \frac{1}{L^2}\Delta\theta_1 
\vspace*{0.2cm} \\
	\displaystyle 
\varphi_{\bar q} = \pi - \phi_1 
- \frac{1}{L}\delta\phi_1  - \frac{1}{L^2}\Delta\phi_L
 \end{array}
\right. 
\quad {\rm and} \quad 
 \left\{
 \begin{array}{c}
  \displaystyle 
\vartheta_q =  \pi + \theta_L + \frac{1}{L}\delta\theta_L 
+ \frac{1}{L^2}\Delta\theta_L 
\vspace*{0.2cm}\\
	\displaystyle 
\varphi_q = \pi - \phi_L 
- \frac{1}{L}\delta\phi_L - \frac{1}{L^2}\Delta\phi_L
 \end{array}
\right. 
\,. \label{relax}
\end{eqnarray} 
In general $\la {\cal L} | \Gamma^{\rm defect-L} | {\cal L} \ra$ is
expanded in terms of $1/L$ as
\begin{eqnarray}
\la {\cal L} | \Gamma^{\rm defect-L} | {\cal L}\ra 
= \lambda\left\{a_0 + a_1\frac{1}{L} + a_2 \frac{1}{L^2} 
\right\} + \mathcal{O}\left(\frac{1}{L^3}\right)\,,  
\end{eqnarray}
but under the conditions (\ref{relax}) the constants $a_0$ and $a_1$
vanish. In other words, the leading parts of (\ref{relax}) are 
determined so that we can take a regular BMN limit. 
Then the terms with higher order than $1/L^3$ also vanish
in the $L\rightarrow\infty$ limit. 
The second-order term gives only a non-vanishing contribution 
\begin{eqnarray}
  \langle {\cal L} | \Gamma^{\rm defect-L} | {\cal L} \rangle
&=&
 \frac{\lambda}{8 \pi^2} \times \frac{1}{2L^2} \bigg\{ (\delta\theta_1)^2 
+ \sin^2 \theta_1 (\delta\phi_1)^2 \bigg\} \\
&=& \frac{1}{4 \pi^2} \frac{\tilde{\lambda}}{2} 
\bigg\{ |\delta \mbox{\boldmath{$m$}}_1|^2 - |\mbox{\boldmath{$m$}}^*_1 \cdot
\delta \mbox{\boldmath{$m$}}_1 |^2 \bigg\}\,.  
\label{HamiltLeft}
\end{eqnarray} 
This will be absorbed into the action obtained from the bulk
contribution. The result that the non-diagonal integrable boundary
terms give the same action as the bulk part would be rather
non-trivial, though it is physically quite natural.  
The same argument can be applied for $\langle {\cal R} |
\Gamma^{\rm defect-R} | {\cal R} \rangle $\,.

The Dirichlet and Neumann boundary conditions for open strings 
may be encoded into $\delta m_1$ in
(\ref{HamiltLeft})\,.  Those cannot be however determined before taking
the continuum limit because the $\sigma$-dependence of 
\mbox{\boldmath{$m$}} should be explicitly specified. 
It is necessary to take a variation of the continuum action 
in order to determine the boundary conditions. 

Finally we shall give a simple geometrical interpretation of the
conditions $a_0 = a_1=0$\,. 
We focus on the boundary on the left hand
side since the analysis on the other side is similar. The key point is
that $n_0^K \equiv \langle\bar{q} | \sigma^K | \bar{q}\rangle $ should
be interpreted as the coordinates for the end-point of the open string
on S$^2$ \cite{Open-sol}. Since $(L^K)_{I_1 J_1} \equiv i
\epsilon_{I_1KJ_1}$ are the $SO(3)_{\rm H}$ generators, $n_1^K \equiv
\langle m_1 | L^K | m_1 \rangle $ should denote the nearest point from
the end-point. Then we find that $a_0=0$ leads to 
$1 - \mbox{\boldmath $n$}_0 \cdot \mbox{\boldmath $n$}_1 =0\,.$  
In the continuum limit, $\mbox{\boldmath $n$}_0$ and $\mbox{\boldmath
$n$}_1$ are very close and hence we can expand as $\mbox{\boldmath
$n$}_1 \simeq \mbox{\boldmath $n$}_0 + d\mbox{\boldmath $n$}_0$\,. Then
the condition $a_1=0$ leads to 
$ \mbox{\boldmath $n$}_0 \cdot d\mbox{\boldmath $n$}_0 = 0\,.$ 
Thus the conditions $a_0=a_1=0$ imply that the endpoints of the open
string should lie on the S$^2$-brane.

\subsection{Derivation of Landau-Lifshitz type Sigma Model from the dCFT}

Now let us consider the expectation value of the open spin chain
Hamiltonian in terms of coherent states (\ref{m-coh}), (\ref{q-coh}). 

The expectation value of $D_{SO(6)}$ in the whole coherent states 
$|m \ra$ consists of three parts as 
\begin{eqnarray}
\la m | D_{SO(6)} | m \ra &=& \la m |\Gamma^{\rm bulk}_{\cal O} | m \ra 
+ \langle{\cal L}|\Gamma_{\cal O}^{\rm defect-L} | {\cal L} \rangle
+ \langle{\cal R}|\Gamma_{\cal O}^{\rm defect-R} | {\cal R} \rangle\,. 
\label{EVD6}
\end{eqnarray} 
Here the defect contributions have been already evaluated and the next
task is to evaluate the bulk contribution. Following the procedure in
\cite{ST-coh}, we shall introduce an antisymmetric $3 \times 3$ matrix
$m^{IJ}$ defined as 
\begin{eqnarray}
 m^{IJ}
\equiv
 \langle m | M^{IJ} | m \rangle = m^a{}^* M^{IJ}_{ab} m^b =  m^I{}^* {m^J} 
- m^I m^J{}^* \quad (I,J=1,2,3)\,. \label{mij}
\end{eqnarray} 
The expectation value of $H_{l,l+1}$ is evaluated as
\begin{eqnarray}
 \la m | H_{l,l+1} | m \ra &=&
 \frac{15}{16}  \sum_{I,J=1}^3 m_l^{IJ} m_{l+1}^{IJ} 
- \frac{1}{16}\sum_{I,J,K,L=1}^3 m_l^{IJ} m_l^{JK} m_{l+1}^{KL} m_{l+1}^{LI}
\nn \\
&=& \frac{15}{32} \mbox{Tr}(m_l - m_{l+1} )^2
 + \frac{1}{32} \mbox{Tr}[(m_l^2 - m_{l+1}^2)^2]\,.
\end{eqnarray} 

Then we consider a continuum limit $L\to \infty$ with $\tilde{\lambda}
\equiv \frac{\lambda}{L^2}$ fixed. 
We expand in powers of $\pi/L$ as
\begin{eqnarray}
 m^{IJ}(\sigma_{l+1}) = m^{IJ}(\sigma_l) + \frac{\pi}{L}\partial_\sigma 
m^{IJ} + \cdots\,, 
\end{eqnarray}
and keep the leading terms. The continuum limit is taken as 
\begin{eqnarray}
 \langle m | D_{SO(6)} | m \rangle
&\to& 
 \frac{\lambda}{16 \pi^2} L \int_0^\pi\!\!\frac{d\sigma}{\pi} 
 \left(\frac{\pi}{L}\right)^2 
 \mbox{Tr}\left[
 \frac{15}{32}(\partial_\sigma m)^2 
 + \frac{1}{16} m^2 (\partial_\sigma m)^2 + \frac{1}{16} 
(m \partial_\sigma m)^2\right] \nn \\
&\simeq& 
 L \int_0^\pi\!\!\frac{d \sigma}{\pi}\, \frac{\tilde{\lambda}}{32} 
\mbox{Tr}\left[ 
(\partial_\sigma m)^2 + \frac{1}{16}(m \partial_\sigma m)^2\right]\,,   
\label{EV6-2}
\end{eqnarray} 
where we have used $\mbox{Tr}(\partial_\sigma m)^2 = 2
\mbox{Tr}\left[m^2(\partial_\sigma m)^2\right] 
+ \mbox{Tr}\left(m \partial_\sigma m\right)^2$\,. 
By substituting (\ref{mij}) into (\ref{EV6-2}),
\begin{eqnarray} 
 \langle m | D_{SO(6)} | m \rangle 
&\simeq& 
 \frac{1}{4} L \int_0^\pi\!\! \frac{d\sigma}{\pi}\,
 \frac{\tilde{\lambda}}{2} \Bigl\{
|\partial_\sigma \mbox{\boldmath $m$}|^2 -  
|\mbox{\boldmath $m$}{^*} \cdot \partial_\sigma \mbox{\boldmath $m$}|^2
\Bigr\} \nn \\
&=& - L \int_0^\pi\!\!\frac{d\sigma}{\pi}
\Bigl\{- \frac{1}{2}\cdot\frac{\tilde{\lambda}}{4} | D_\sigma \mbox{\boldmath $m$}|^2 \Bigr\} 
\equiv -L \int_0^\pi\!\!\frac{d\sigma}{\pi} 
\langle m |{\cal H}| m\rangle\,. \label{EV6-3}
\end{eqnarray}
The phase-space Lagrangian $\mathcal{L}$ is defined as 
\begin{eqnarray}
 \mathcal{L} \equiv - \la m | i\frac{d}{d t}|m \ra + \la m|{\cal H}|m\ra\,,  
\end{eqnarray}
and the resulting sigma model action is given by 
\begin{eqnarray}
I = L \int\!\! dt\! \int_0^\pi\!\!
\frac{d\sigma}{\pi}\, {\cal L}\,, \qquad {\cal L} = 
-i \mbox{\boldmath $m$}^* \cdot \partial_t \mbox{\boldmath $m$}
 - \frac{1}{2}\cdot\frac{\tilde{\lambda}}{4} 
|D_\sigma \mbox{\boldmath $m$}|^2
\,. \label{sigmaSYM}
\label{sym}
\end{eqnarray} 
In the next section we will reproduce the same result as (\ref{sigmaSYM}) 
from the string side.

\subsubsection*{Comment on the $SO(6)/SO(5)$ case} 

We may consider the case 2 related to $H=SO(5)$ in the unbroken $SO(6)$
case. When we follow the above analysis in this case, we cannot
completely delete the contribution of the additional anomalous dimension
because the $\epsilon_{IJK}$ terms vanish due to the reality of the
$SO(3)_{\rm H}$ coherent state (i.e. $\la m|_{I}\epsilon_{IJK} |m\ra_{K}
= 0$). Thus the Kronecker delta terms still remain even after taking the
expectation value of the spin chain Hamiltonian, and lead to the term
proportional to $\lambda$\,. This contribution is absorbed into the
large ``extensive'' one-loop shift that comes from the bulk
contribution, but the problem of the consistent continuum limit still
remains. After all, the problem of the defect interactions reduces to
that of the continuum limit.

\section{Landau-Lifshitz Type Sigma Model from the String Action}

\subsection{Derivation of Landau-Lifshitz Type Sigma Model}

We shall derive a Landau-Lifshitz type sigma model action as
an effective action of the open string moving almost at the speed of
light, following \cite{KT-coh}. In our case, open strings can move only
on the S$^2$ spanned by $X_{\rm H}^I\ (I=1,2,3)$ with
$\sum_{I=1}^3(X_{\rm H}^I)^2=1$. Hence  
we start from the bosonic part of the phase-space Lagrangian for an open
string on $R\times$S$^2$\,,  
\begin{equation} \label{lak} 
{\cal L} = -\frac{1}{2}\kappa^2 + P_{\rm H}^I \dot{X_{\rm H}^{I}} 
- \frac{1}{2}P_{\rm H}^I P_{\rm H}^I 
- \frac{1}{2} {X_{\rm H}^I}{}' {X_{\rm H}^I}{}' - \frac{1}{2}
\Lambda (X_{\rm H}^I X_{\rm H}^I -1 )\,, 
\end{equation}
where the symbols ``$\cdot$'' and ``{\footnotesize $\prime$}'' mean the
derivatives with respect to the world-sheet coordinates $\tau$ and
$\sigma$, respectively. Here we have chosen the conformal gauge and the 
additional gauge-fixing condition $t=\kappa \tau$\,. 

Next we consider an effective action where the slowly-changing
coordinates only survive, and so we first isolate a fast coordinate. In
this process, it is inevitable to use the phase-space description.  Then
we take a limit in which the velocity of the fast coordinate is put to
infinity. Finally we get the following Landau-Lifshitz type effective
action
\begin{eqnarray} 
\label{vbvb} && I = L \int\! dt
\!\int_0^{\pi}\!\frac{d\sigma}{\pi}\,{\cal L}\,, \quad {\cal L} = - i
V^* \dot{V} - \frac{1}{2} |D_\sigma V|^2\,, 
\end{eqnarray} 
where $L= \sqrt{\lambda}\,l~(l\simeq \kappa)$ and the symbol ``$\cdot$''
implies the
derivative with respect to the time $t$, instead of $\tau$\,. The time
coordinate is rescaled as $t \rightarrow \kappa^2 t$ and the Lagrange
multiplier terms are omitted. In addition, when we rescale the time $t$
as $t \rightarrow \frac{1}{4}\tilde{\lambda}\,t$\,,\footnote{The factor
$1/4$ arises since $\tilde{\lambda}\equiv \lambda/L^2$ and 
the $L$ for closed strings is replaced by $2L$ for open
strings. } the Wess-Zumino term is kept invariant but the remaining part
is shifted as 
\begin{eqnarray}
I = L \int\! dt \!\int_0^{\pi}\!\frac{d\sigma}{\pi}\,{\cal L}\,, \quad
{\cal L} = - i V^* \dot{V} - \frac{1}{2}\cdot\frac{\tilde{\lambda}}{4}
|D_\sigma V|^2\,.
\label{5.10}
\end{eqnarray} 
We thus find a complete agreement of the sigma model actions (\ref{sym})
in SYM and (\ref{5.10}) in the string side, under the identification
of {\boldmath{$m$}} with $V$\,. 

In the closed string case, the Landau-Lifshitz type sigma model can also 
be derived from the approach of Mikhailov \cite{Mikhailov}.  It would
be an interesting practice to rederive the action in our open string
case, by following \cite{Mikhailov}.

\subsection{$SO(3)_{\rm H}$ Pulsating String Solution}

We can find a pulsating solution under a solution ansatz. Let us
parametrize the complex vector $V^I$ as 
\begin{equation}
V^I = \frac{a^I - ib^I}{\sqrt{2}}\,, \quad 
V^I {V^I}^* =1\,, \quad (V^I)^2 =0\,. 
\end{equation}
By setting the real vectors $a$ and $b$ as
\begin{eqnarray}
\label{ansatz} 
a = (\cos\th \cos\phi, -\cos\th\sin\phi,\sin\th)\,, \quad 
b = (- \sin\phi, -\cos\phi,0)\,,
\end{eqnarray}
the Lagrangian becomes 
\begin{eqnarray}
\mathcal{L} = \cos\th\, \dot{\phi} - \frac{\tilde{\lambda}}{16}
\left({\th'}^2 + \sin^2\th\, {\phi'}^2\right)\,,
\end{eqnarray}
in terms of two coordinates $\th$ and $\phi$ of S$^2$\,. The
pulsating solution is now described by taking a special solution $\th =
\pi/2$ and $\phi = m\sigma$ where $m$ is an integer\footnote{For more general pulsating string solutions, see \cite{EMZ,pulsating3}.}. The energy of the
pulsating string solution is obtained as
\begin{eqnarray}
E = \frac{1}{16}\tilde{\lambda}L\sin^2\th\,{\phi'}^2 
= \frac{\lambda}{16L}m^2\,. 
\end{eqnarray}
It completely agrees with the result of \cite{pulsating1} in the open
string case. Here we have utilized the doubling trick formula that is
discussed by Stefanski \cite{Open-sol},
\begin{eqnarray}
E_{\rm open}(L) = \frac{1}{2}E_{\rm closed}(2L)\,. 
\label{doubling}
\end{eqnarray}
Then the energy for the pulsating closed string solution is 
\begin{eqnarray}
E_{\rm closed} =  \frac{\lambda}{4L}m^2\,,
\end{eqnarray} 
and hence the doubling trick formula (\ref{doubling}) is surely
satisfied. 

We should remark that the pulsating closed string solution is also a
solution in our open string case by imposing Neumann and Dirichlet
conditions at the end-points: 
\begin{eqnarray}
\mbox{N}: \quad \partial_{\sigma}V(t,\sigma=0,\pi) = 0\,, \qquad 
\mbox{D}: \quad V(t,\sigma=0,\pi) = 0\,. 
\label{bound} 
\end{eqnarray}  
In the parametrization of (\ref{ansatz}) and the solution ansatz:
$\th=\pi/2$ and $\phi = m\sigma$\,, $V^2$ and $V^3$ satisfy Neumann
conditions, and $V^1$ does a Dirichlet one.

\subsection{Construction of Coherent Defect Operators}

Finally we shall comment on the forms of defect operators corresponding
to open string solutions. The defect operators naturally associated to
open semiclassical string solutions should be ``locally BPS'' coherent
defect operators. Slightly generalizing the factorization ansatz in the closed
string case \cite{KT-coh}, we suppose that the coefficient of
(\ref{cop}) is decomposed as
\begin{eqnarray}
 \psi_{m I_1 I_2 \cdots I_L n} = \eta_m' m_1^{I_1} m_2^{I_2} \cdots 
m_L^{I_L} \eta_n\,.
\end{eqnarray} 
Here $m_l$'s are the coherent states (\ref{m-coh}) and correspond to
open string solutions under the identification of $\mbox{\boldmath $m$}$
with $V$ in the continuum limit. 
By assuming that $\eta'_m$ and $\eta_n$ are given as 
$ (\eta_m') = |\bar{q} \rangle \quad \mbox{and} \quad (\eta_n) =
|q\ra$\,, the corresponding operators are 
\begin{eqnarray}
 {\cal O} = [\eta_m' \bar{q}_m] \left[\prod_{l=1}^L m_l^{I_l} X_{\rm H}^{I_l}
\right] [\eta_n q_n]\,. \label{localBPS}
\end{eqnarray} 
In particular, when we set $\theta_l=\phi_l=0$ for all $l$\,, 
\begin{eqnarray}
 (\eta_m') = \left[
	\begin{array}{@{\,}c@{\,}}
	 1 \\
	 0
  \end{array} 
 \right]\,,
\quad 
 (\eta_n) =
 \left[
	\begin{array}{@{\,}c@{\,}}
	 0 \\
	 1
  \end{array} 
 \right]\,,
\qquad 
 (m_l^{I_l})
=
 \left[
 \begin{array}{@{\,}c@{\,}}
  1 \\
  i \\
  0
 \end{array} 
 \right]
\quad (l=1,\cdots, L)\,, 
\end{eqnarray} 
we can consistently reproduce the open string BPS vacuum \cite{LP} 
\begin{eqnarray}
 {\cal O} = \bar{q}_1 Z^L q_2 
\qquad ( Z \equiv X_{\rm H}^1 + i X_{\rm H}^2)\,. 
\end{eqnarray}

In addition we may construct the operator corresponding to the pulsating
open string solution discussed above by putting the following data into
(\ref{localBPS})\,,
\begin{eqnarray}
 (\eta_m') =
 \frac{i}{\sqrt{2}}
 \left[
	\begin{array}{@{\,}c@{\,}}
	 1 \\
	 -1
  \end{array} 
 \right]\,,
\quad 
 (\eta_n)
=
  \frac{i^{m+1}}{\sqrt{2}}
 \left[
	\begin{array}{@{\,}c@{\,}}
	 1 \\
	 (-1)^m
  \end{array}
 \right]\,,
\quad 
 (m_l^{I_l})
=
 \frac{e^{-im\sigma_l}}{\sqrt{2}}
 \left[
 \begin{array}{@{\,}c@{\,}}
   i \sin(m \sigma_l) \\
   i \cos(m \sigma_l)\\
   1
 \end{array}
 \right]\,.
\end{eqnarray} 
Here a few examples have been discussed, but we need more efforts to
confirm the ansatz (\ref{localBPS})\,.

\section{Other AdS-brane Cases}
\label{general:app}

We have discussed the case that an AdS$_4\times$S$^2$-brane is inserted
in the AdS$_5\times$S$^5$ background. By considering other AdS-branes,
one may obtain other defect conformal field theories (For the
classification of possible AdS-branes, see the work of Skenderis and
Taylor \cite{ST} or an approach from $\kappa$-symmetry
\cite{SY:string}). The possible 1/2 supersymmetric AdS-branes are
summarized in Tab.\,\ref{brane:tab}. For all cases, the holographic
duals are described by dCFTs. The feature of the defect fields is
different for the number of Neumann-Dirichlet directions ($\sharp$ND).
In the case of $\sharp$ND=4 intersections, the ``quark'' field $q$
supplied by the defect is a complex scalar field, 
while $q$ is a fermion field in the case of $\sharp$ND=8
intersections.

\begin{table}[htbp]
\begin{center} 
\begin{tabular}{cc}
\begin{tabular}{|c|c|c|}
\hline 
Brane & $\sharp$ND=4 & Embedding \\
\hline\hline 
D1 & (0$|$ D1 $\bot$ D3) & AdS$_{2}$ \\
\hline 
D3 & (1$|$ D3 $\bot$ D3) & AdS$_3\times$S$^1$ \\
\hline  
D5 & (2$|$ D5 $\bot$ D3) & AdS$_{4}\times$S$^{2}$ \\
\hline 
D7 & (3$|$ D7 $\bot$ D3) & AdS$_{5}\times$S$^{3}$ \\
\hline
\end{tabular} 
\quad & \quad  
\begin{tabular}{|c|c|c|} 
\hline 
Brane & $\sharp$ND=8 &  Embedding \\
\hline\hline 
D5 & (0$|$ D5 $\bot$ D3) & AdS$_2\times$S$^4$ \\
\hline 
D7 & (1$|$ D7 $\bot$ D3) & AdS$_3\times$S$^5$ \\
\hline
  \end{tabular} 
\end{tabular} 
\end{center}
\caption{The possible configurations of 1/2 supersymmetric AdS-branes. }
\label{brane:tab}
\end{table}

When we focus upon the analysis of the S$^5$ part, open strings can live
on the S$^n$ part $(n=2,3,4,5)$\footnote{The $n=0$ case has no part in
S$^5$\,. For the $n=1$ case, the group $SO(2)_{\rm H}$ is abelian and the $H$
becomes trivial. Then the coherent state leads to the BPS vacuum.  Hence
we will not consider them here.} that is a part of the above-mentioned
AdS-branes.  In the case of the S$^n$\,, we would need the $SO(n+1)_{\rm
H}$ coherent states to evaluate the classical sigma model action by
taking a continuum limit. We argue that the appropriate choice of the
maximal stability subgroup $H$ and the vacuum $|0\ra$ would be
\begin{eqnarray}
H = SO(n-2) \times SO(2)\,, \qquad |0\ra = 
(\underbrace{0,\ldots,0}_{n-1},1,i)\,. 
\end{eqnarray}
That is, the coherent states for general AdS-branes are constructed for
the coset:
\begin{eqnarray}
\label{coset}
SO(n+1)_{\rm H}/[SO(n-2) \times SO(2)]\,. 
\end{eqnarray}
By using the coherent state for the coset (\ref{coset}),  
\begin{eqnarray}
|m\ra = \exp\left[\sum_{i=1}^{n-1}(a_iM_{i5} + a_{i+4}M_{i6})\right]|0\ra\,,
\end{eqnarray}
with $|0\ra = (0,\ldots,0,1,i)$\,, we can evaluate the expectation value
of the integrable open spin chain Hamiltonian $D_{SO(6)}$\,. Following
the work of Stefanski and Tseytlin \cite{ST-coh}, let us introduce an
antisymmetric imaginary $(n+1) \times (n+1)$ matrix $m^{ij}\equiv \la m
| M^{ij}|m \ra~$\,. The only difference from the work \cite{ST-coh} is
the range of the index $i$\,.  
The integrable
boundary terms may possibly vanish according to our description and then
the expectation value of $H_{l,l+1}$ is evaluated as follows:
\begin{eqnarray}
\la m|H_{l,l+1}|m \ra &=& \frac{1}{2}\left\{1 -
\frac{1}{16}(n-1)\right\}{\rm Tr}(m_l - m_{l+1})^2 +
\frac{1}{32}(n-1){\rm Tr}(m_l^{2} - m_{l+1}^2)^2\,.
\end{eqnarray}
After taking a continuum limit, we obtain the following
classical sigma model action:
\begin{eqnarray}
\la m| D_{SO(6)}|m\ra \rightarrow 
L\int^{\pi}_0\!\!\frac{d\sigma}{\pi}\,\frac{\tilde{\lambda}}{32}
\left[
{\rm Tr}(\partial_{\sigma}m)^2 + \frac{1}{16}(n-1)
{\rm Tr}(m\partial_{\sigma}m)^2
\right]\,, 
\end{eqnarray}
where we have used the formula: 
\begin{eqnarray}
{\rm Tr}[m^2(\partial_\sigma m)^2] = 
\frac{1}{2}{\rm Tr}(\partial_{\sigma}m)^2 - \frac{1}{2}{\rm Tr}
(m\partial_{\sigma}m)^2\,. 
\end{eqnarray}
Here we should notice the identity ${\rm Tr}(m\partial_{\sigma}m)^2 =0$
and thus the resulting sigma model action is independent of $n$\,.  That
is, the $n$-dependence appears at the quantum spin chain level but it
disappears after taking the continuum limit. As the result, by
introducing the complex unit vector $V^i$, we obtain the sigma model
Lagrangian, 
\begin{eqnarray}
\mathcal{L} &=& -iV^{i\ast}\partial_{t}V^i 
- \frac{1}{2}|D_{\sigma}V^i|^2\,. \label{sigma}
\end{eqnarray}
This expression is identical for each of AdS-branes,
except the range of the index $i$\,. 

The sigma model action (\ref{sigma}) can be rederived from the string
action in the same way as in section 5. Thus it may be argued that the
complete agreement of the sigma model action would be valid for general
AdS-branes as well as AdS$_4\times$S$^2$\,.  This fact would ensure that
the supersymmetric AdS-branes may be included in studies of the
correspondence between semiclassical strings and SYM operators.

Moreover, our consideration for coherent states may be generalized
for some cases other than AdS-branes, for example, orbifold
backgrounds \cite{BGMNN,Open-sol,CWW,Ideguchi}. A study in this direction
is favorable and interesting.

\section{Conclusion and Discussion}

We have discussed the correspondence between long defect operators in
the dCFT and open semiclassical strings. The coherent states for the
integral open spin chain have been constructed. In particular, we have
given a prescription to treat the boundary terms in the open spin chain
Hamiltonian. By evaluating the expectation value of the Hamiltonian with
the coherent states and taking the continuum limit, a Landau-Lifshitz
(LL) type of sigma model has been obtained. This action has been also
derived by starting from the string action with the appropriate gauge
fixing and $\kappa\rightarrow\infty$\,.  In conclusion, for an open
string case, we have found a complete agreement of the sigma model
actions in both sides as in closed string cases. It has been also found
that an $SO(3)_{\rm H}$ pulsating solution is included as a solution of
the derived LL-type sigma model and its energy is completely identical
to the result obtained in a different method \cite{pulsating1}. In
addition, we have discussed the sigma model action in other general
AdS-brane cases. When we consider other AdS-branes, the index range of
the sigma model variable is different in each of cases according to
S$^n$ $\subset$ S$^5$ ($n=2,\ldots,5$). However, we have seen that this
difference may not affect on the resulting sigma model action and thus
we argue that the similar LL-type sigma model would be obtained in other
AdS-brane cases. 

We have discussed the one-loop dilatation operator for defect operators
composed of the $SO(6)$ scalar fields. It is an interesting future work
to investigate at two-loop level. The $SO(6)$ sector is closed at
one-loop level, but it is not the case at two-loop level. That is, the
fields belonging to other sectors may mix into the $SO(6)$ scalar
sector. The mixing of fermions occurs at two-loop level, but recently an
interesting possibility has been proposed by Minahan
\cite{Minahan:higher}. Following the idea of \cite{Minahan:higher}, it
may be suppressed in taking $L\rightarrow\infty$ limit. Thus, the
two-loop integrability would possibly hold in this limit.  It is
important to calculate the two-loop defect interaction in the system we
considered and check the consistency to the doubling trick in the string
side at two-loop level. We will work in this direction and report the
result in the near future \cite{future}.

It is also an interesting subject to clarify the connection between the
integrability and supersymmetric D-branes (i.e., BPS conditions). The
relation between BPS conditions and rotating string solutions is
discussed in \cite{MMT,BCM}. A study in this direction would be helpful
to investigate such an issue. In addition, it is interesting to clarify
the relationship of the integrabilities (Yangian symmetries) of the
Landau-Lifshitz type sigma models in closed and open cases (For a closed
case, see \cite{AR}).

\section*{Acknowledgments}

We would like to thank H.~Fuji, M.~Hatsuda, H.~Kawamura, D.~Tomino and
A.~Yamaguchi for useful discussion.  The work of K.~Y.\ is supported in
part by JSPS Research Fellowships for Young Scientists.

\end{document}